\documentclass[twocolumn,aps,prd,epsf,showpacs]{revtex4}

\usepackage[dvips]{graphicx}
\usepackage{epsfig}
\usepackage{color}
\usepackage{amsmath}
\usepackage[utf8]{inputenc}
\usepackage[english]{babel}

\newcommand{\be}{\begin{equation}}
\newcommand{\ee}{\end{equation}}
\newcommand{\bea}{\begin{eqnarray}}
\newcommand{\eea}{\end{eqnarray}}

\newcommand{\gtapprox}{\raisebox{-0.5ex}{$\,\stackrel{>}{\scriptstyle\sim}\,$}}
\newcommand{\ltapprox}{\raisebox{-0.5ex}{$\,\stackrel{<}{\scriptstyle\sim}\,$}}

\begin{document}


\title{Lattice QCD signal for a bottom-bottom tetraquark}

\author{Pedro Bicudo}
\email{bicudo@ist.utl.pt}
\affiliation{Dep.\ F\'{\i}sica and CFTP, Instituto Superior T\'ecnico, Av.\ Rovisco Pais, 1049-001 Lisboa, Portugal}

\author{Marc Wagner}
\email{mwagner@th.physik.uni-frankfurt.de}
\affiliation{Johann Wolfgang Goethe-Universit\"at Frankfurt am Main, Institut f\"ur Theoretische Physik, Max-von-Laue-Stra{\ss}e 1, D-60438 Frankfurt am Main, Germany}
\affiliation{European Twisted Mass Collaboration (ETMC)}

\begin{abstract}
Utilizing lattice QCD results for the potential of two static antiquarks and two dynamical quarks as well as quark model techniques for the dynamics of two heavy antiquarks in a cloud of two light quarks, we are provided with an accurate framework for the study of possibly existing heavy-heavy-light-light tetraquarks. Among the possible quantum numbers of such a system, we find binding in only one channel, the scalar isosinglet. Solving the Schr\"odinger equation for the displacement of the heavy antiquarks and taking systematic errors into account, we find an antibottom-antibottom-light-light bound state with a confidence level of around $1.8 \sigma \ldots 3.0 \sigma$ and binding energy of approximately $30 \, \textrm{MeV} \ldots 57 \, \textrm{MeV}$.
\end{abstract}

\pacs{12.38.Gc, 13.75.Lb, 14.40.Rt, 14.65.Fy.}

\maketitle


\section{Introduction}

Experimentally exotic hadrons have been searched for many years, because as soon as quarks were proposed in the sixties, it became clear that systems more complex than mesons and baryons could possibly exist. However, exotic hadrons are very elusive systems. They are much harder to observe experimentally, to understand by theoretical model calculations, and to simulate by means of lattice QCD than the conventional mesons and baryons. To confirm the existence or non-existence of exotic hadrons still remains an important problem in QCD. 

A frequently discussed exotic multiquark is the tetraquark. It was already proposed in the seventies \cite{Jaffe:1976ig} as a bound state formed by two quarks and two antiquarks. There are several hadronic resonances, which are tetraquark candidates, e.g.\ $\sigma$, $\kappa$, $D_{s0}^\ast$ or $D_{s1}$ \cite{PDG}. The most recent tetraquark candidate has been claimed by the BELLE Collaboration \cite{Belle:2011aa}, observing in five different $\Upsilon(5S)$ decay channels two new charged bottomonium resonances $Z_b$ with masses $10610 \, \textrm{MeV}$ and $10650 \, \textrm{MeV}$ and narrow widths of the order of $15 \, \textrm{MeV}$, where the charge can only come form the presence of a light quark and a light antiquark. However, the tetraquark nature of these resonance is disputed \cite{Bugg:2011jr}.

Notice that the experimental particle physics collaborations are technically improving rather impressively. One decade ago SELEX at FNAL already studied the doubly charmed baryons. Presently, BELLE at KEK, CDF and D$\O$ at FNAL, and LHCb at CERN have already observed bottom hadrons. Thus, they may possibly be able to search, not only for bottom-antibottom tetraquarks, but also for  bottom-bottom tetraquarks, in case sufficiently strong evidence is presented by theoretical calculations.

Tetraquark studies face several difficulties. (1) Mesons and baryons only decay strongly when the confining string breaks, a quark antiquark pair is created and and either two mesons or a meson and a baryon are formed. In contrast to that, tetraquarks are directly open to meson-meson decay. (2) Moreover, tetraquarks are relativistic four-body systems, which are highly complex few-body systems. (3) And on the top of these technical difficulties no model (since the onset of QCD \cite{Jaffe:1976ig} up to the present) seems to be sufficiently well calibrated to address multiquark binding: different quark models, even when producing similar meson or baryon spectra, usually differ significantly in predictions for tetraquarks. Exceptions are heavy four-quark systems, say an exotic $c c \bar{b} \bar{b}$, but these are extremely hard to investigate experimentally.

An example of a rather complex system to study in lattice QCD is a tetraquark including a $b$ and a $\bar{b}$ quark. It is very interesting, due to the BELLE observation \cite{Belle:2011aa} of $B^\ast \bar{B}$ and $B^\ast \bar{B}^\ast$ tetraquark candidates. However, it couples to at least five decay channels as reported by BELLE. Therefore, we study here in a first step the theoretically simpler $B B$ system. In the near future we plan to extend our investigations to the $B \bar{B}$ tetraquark. Notice that the observation of a $B \bar{B}$ system at BELLE suggests that a $B B$ tetraquark may also be observable in present day laboratories.

A strategy to avoid many technical difficulties associated with tetraquark studies consists in searching for bound states rather than for resonances, which is e.g.\ appropriate, when two heavy antiquarks (or equivalently two heavy quarks) are involved. This strategy was already identified in the eighties \cite{Ballot:1983iv}. On the one hand, it is plausible that any light tetraquark can only be treated as a resonance, because it couples to two-meson channels with identical quantum numbers, where at least one of the mesons is a pion or kaon. Due to chiral symmetry, these are very light mesons and, consequently, the tetraquark will have a rather light open decay channel. On the other hand the presence of two heavy antiquarks is sufficient to force all involved hadrons, i.e.\ the tetraquark and any two-meson channels with the same quantum numbers, to escape chiral symmetry. Moreover, the heavy antiquarks also reduce the technical complexity of the four quark system, since heavy antiquarks are non-relativistic. This in turn allows for the Born-Oppenheimer approximation: for the light quarks the heavy antiquarks can be regarded as static color charges; once the energy of the light quarks is determined, it can be utilized as an effective potential for the heavy antiquarks.


\section{Heavy antiquark-antiquark interaction}

\subsection{Lattice QCD results}

The major theoretical problem remaining is, to obtain the correct effective potential, which has been studied by lattice QCD methods mainly in the quenched approximation (cf.\ e.g.\ \cite{Stewart:1998hk,Michael:1999nq,Cook:2002am,Doi:2006kx,Detmold:2007wk}). Only recently computations of this potential with dynamical sea quarks have been performed \cite{Wagner:2010ad,Bali:2010xa,Wagner:2011ev}.


Here we use such dynamical results obtained with a comprehensive set of four quark operators of the form
\be
\label{EQN003} (\mathcal{C} \Gamma)_{A B} \Big(\bar{Q}_C(\mathbf{r}_1) \psi_A^{(1)}(\mathbf{r}_1)\Big) \Big(\bar{Q}_C(\mathbf{r}_2) \psi_B^{(2)}(\mathbf{r}_2)\Big) ,
\ee
where $\bar{Q}$ denotes a static quark operator, $\psi$ a light antiquark operator, $A$, $B$ and $C$ are spin indices and $\mathcal{C} = \gamma_0 \gamma_2$ is the charge conjugation matrix \cite{Wagner:2010ad,Wagner:2011ev}. While for the static antiquarks the only relevant variable is their separation, the two light $u/d$ quarks can be combined in $8 \times 8 = 64$ different ways via the $4 \times 4$ matrix $\Gamma$ and the light quark flavors $\psi^{(1)} \psi^{(2)} \in \{ ud - du , uu, ud + du , dd \}$ (each light quark has two isospin, two spin and two parity degrees of freedom). Symmetries and quantum numbers of such four-quark states are explained in detail in \cite{Wagner:2010ad}.

\begin{figure}[h]
\includegraphics[scale=1.0]{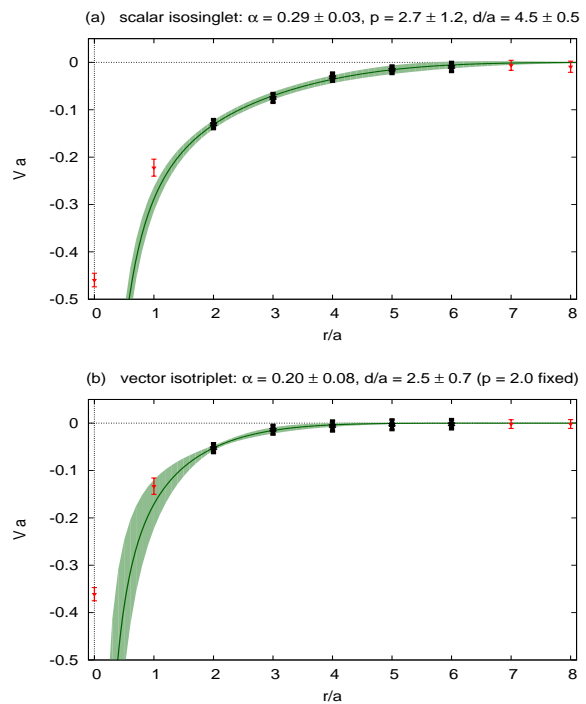}
\caption{\label{FIG002}(Color online). The static antiquark-antiquark potential as a function of the separation in units of the lattice spacing $a \approx 0.079 \, \textrm{fm}$. (a)~Three parameter fit ($\alpha$, $d$, $p$) of ansatz (\ref{ansatz}) to the most attractive channel, the scalar isosinglet. (b)~Two parameter fit ($\alpha$, $d$; $p = 2.0$ fixed) of ansatz (\ref{ansatz}) to the less attractive vector isotriplet.}
\end{figure}

In this work we focus on the two attractive channels between ground state static-light mesons ($B$ and $B^\ast$ mesons): a scalar isosinglet (cf.\ Fig.\ \ref{FIG002}(a)) with corresponding four-quark creation operator
\be
\psi^{(1)} \psi^{(2)} = ud - du \quad , \quad \Gamma = \gamma_5 + \gamma_0 \gamma_5 ,
\ee
which is most attractive, and a vector isotriplet (cf.\ Fig.\ \ref{FIG002}(b)) with corresponding four-quark creation operator e.g.\
\be
\psi^{(1)} \psi^{(2)} = ud + du \quad , \quad \Gamma = \gamma_3 + \gamma_0 \gamma_3 ,
\ee
which is less attractive. Note that these operators have not only specific quantum numbers, but also exhibit a structure particularly suited, to excite the ground state (a system composed at large $b \bar{b}$ separations of pseudoscalar $B$ and/or vector $B^*$ mesons, but not of excited positive parity mesons such as $B_0^\ast$ or $B_1^\ast$) as explained in \cite{Wagner:2010ad,Wagner:2011ev}. Consequently, the resulting potential does not depend on the details of the used operators. In particular, no additional assumption about the physical structure of the four-quark state is made or entering the computation. The arrangement of the four quarks is decided by QCD dynamics, i.e.\ automatically realized in the lattice result according to QCD (cf.\ e.g.\ also recent lattice work on tetraquark candidates, where it has been demonstrated that operators similar to (\ref{EQN003}) generate significant overlap to a variety of different four-quark structures including mesonic molecules, diquark-antidiquark pairs or two essentially non-interacting mesons \cite{Daldrop:2012sr,Alexandrou:2012rm,Wagner:2012ay,Wagner:2013nta}).


For further details regarding the lattice computation of the heavy antiquark-antiquark interaction we refer to \cite{Wagner:2010ad,Wagner:2011ev}.


\subsection{\label{SEC003}Screening ansatz}

\begin{figure}[h]
\includegraphics[scale=1.0]{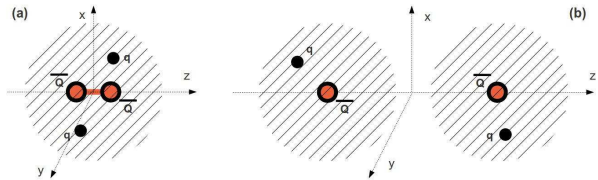}
\caption{\label{strings_screened}(Color online). Screening of the antidiquark flux tube interaction. We show two scenarios: (a) the heavy antiquarks are close and the screening of the light quarks has little effect on the antiquark-antiquark interaction; (b) the heavy quarks are well separated and their color charge is totally screened by the light quark wavefunctions.}
\end{figure}

To motivate an ansatz to fit the lattice results for these potentials, first remember that the pair of heavy antiquarks is immersed in a cloud of two light quarks. The size of this cloud is crucial for the heavy antiquark-antiquark interaction. When the two antiquarks are much closer than twice the typical light quark cloud radius in a heavy-light meson, the antiquark-antiquark interaction is a typical diquark interaction (cf.\ Fig.\ \ref{strings_screened}(a)). However, when the separation of the two antiquarks becomes larger, the light quarks screen the heavy antiquark charges as in Fig.\ \ref{strings_screened}(b) and the antiquark-antiquark interaction vanishes.


The diquark interaction for systems containing only static quarks has also been studied by lattice QCD. Flux tubes have been observed for static baryons, static tetraquarks and static pentaquarks \cite{Cardoso:2011fq,Cardoso:2012rb}. It seems established that at large separations $r$ the diquark potential is linear and confining $\sim \sigma r$. At small separations $r$ the potential is rather Coulomb-like, i.e.\ $\sim -\alpha/r$. A similar system, which has been studied in even more detail by lattice methods is the ``static-static meson'' or static potential; here $\alpha \approx 0.3 \ldots 0.4$, while $\sigma \approx (0.44 \, \textrm{GeV})^2$ is estimated from quark model fits and often used to set the scale. For two heavy antidiquarks in a cloud of two light quarks we expect a similar Coulomb-like potential of order $-\alpha/r$ at small separations. At larger separations (in contrast to the purely static case) the potential should be screened by the light quarks, as discussed in the previous paragraph and illustrated in Fig.\ \ref{strings_screened}(b).

The screening of the heavy color charge is due to the decrease of the wave function $\psi$ of the light quark with respect to its separation from the heavy quark. One expects this decrease to follow an exponential of a power law, i.e.\ $\psi(r) \propto \exp(-(r/d)^p)$, where $d$ characterizes the size of the quark-antiquark system, i.e.\ a $B$ meson. If the quark-antiquark interaction inside a $B$ meson is dominated by a Coulomb-like term, the wave function is similar to that of a hydrogen atom, i.e.\ $p = 1$. In case the potential is rather linear, the non-relativistic Schr\"odinger equation is solved by Airy functions corresponding to $p = 3/2$. A similar, but relativistic treatment of the light quark yields $p = 2$ instead.

The above considerations suggest the following ansatz to model the heavy antiquark-antiquark potential:
\be
\label{ansatz} V(r) = -{\alpha \over r} \exp \left(-\left(r \over d\right)^p\right) ,
\ee
where it is expected that $\alpha \approx 0.3 \ldots 0.4$, $d$ is around half the size of a $B$ meson, i.e.\ $d \ltapprox 0.5 \, \textrm{fm}$, and $p \approx 1.0 \ldots 2.0$.


\subsection{\label{SEC005}Fitting procedure and results}

We perform uncorrelated $\chi^2$ minimizing fits of the ansatz (\ref{ansatz}) to the lattice results for the heavy antiquark-antiquark interactions shown in Fig.\ \ref{FIG002}, i.e.\ we minimize
\be
\label{EQN004} \chi^2 = \sum_{r = 2a,\ldots,6a} \bigg(\frac{V(r) - V^\textrm{lat}(r)}{\Delta V^\textrm{lat}(r)}\bigg)^2
\ee
with respect to the parameters $\alpha$, $d$ and $p$ ($V^\textrm{lat}$ denote the lattice results, $\Delta V^\textrm{lat}$ the corresponding statistical errors). Notice that data points for separations $r/a=0$ and $r/a=1$ are excluded from the fits, because they suffer from lattice discretization errors. For $r/a \geq 2$ it has been checked, that these discretization errors are negligible compared to the statistical errors by using different static quark actions \cite{Wagner:2010ad}. On the other hand, data points for large separations have little relevance since the potential vanishes exponentially fast, due to screening, while the statistical error remains similar. Thus we utilize the points at distances $r/a=2, 3, 4, 5, 6 $ for our fits.

For the scalar isosinglet we are able to determine all three parameters $\alpha$, $d$ and $p$ via fitting. For the vector isotriplet a three parameter fit is not stable; therefore, we only fit two parameters, $\alpha$ and $d$, while fixing the exponent to its expected value $p=2.0$. The fits are also shown in Fig.\ \ref{FIG002}, while numerical results are collected in Table~\ref{fit AEK}. Statistical errors for $\alpha$, $d$ and $p$ have been determined via an elaborate Jackknife analysis starting on the level of the lattice correlation functions. In detail we proceeded as follows.
\begin{itemize}
\item[(1)] From the 480 available samples of the correlation functions of four quark operators (\ref{EQN003}) (corresponding to 480 gauge link configurations) we form 20 essentially independent bins by averaging each time over 24 consecutive gauge link configurations; this binning removes possibly existing correlations in Monte Carlo simulation time.

\item[(2)] From these 20 bins we compute not only the average, but also 20 reduced samples, i.e.\ 20 correlation function averages over 19 of the available 20 bins, each time omitting a different bin.

\item[(3)] On the average and on each reduced sample we compute the heavy antiquark-antiquark potential obtaining $V^\textrm{lat}$ and $V^{\textrm{lat,red},n}$, $n = 1,\ldots,20$, using standard lattice techniques (fitting constants to effective mass plateaus at sufficiently large temporal separations); these results are then used in a standard Jackknife analysis, to obtain a statistical error $\Delta V^\textrm{lat}$.

\item [(4)] The $\chi^2$ minimizing fit of the ansatz (\ref{ansatz}) to lattice potential is not only performed for $V^\textrm{lat}$, but also for the reduced samples $V^\textrm{lat,red,n}$, yielding $(\alpha,d,p)$ and $(\alpha^{\textrm{red},n},d^{\textrm{red},n},p^{\textrm{red},n})$ [or $(\alpha,d)$ and $(\alpha^{\textrm{red},n},d^{\textrm{red},n})$]; as in step (3) a standard Jackknife analysis is used, to obtain statistical errors $(\Delta \alpha,\Delta d,\Delta p)$ [or $(\Delta \alpha,\Delta d)$].
\end{itemize}
The fit of the ansatz (\ref{ansatz}) to the lattice results is uncorrelated, because there are not sufficiently many lattice samples available, to estimate a covariance matrix appropriately. Note, however, that we use the same bins and reduced samples for all temporal separations of the correlation functions $t$ and also for all spatial separations of the heavy antiquarks $r$. Therefore, these bins and reduced samples contain information about possibly existing correlations in $t$ and in $r$, which in turn enters the resulting fit parameters $\alpha$, $d$ and $p$. In other words, although we do not mimimize a correlated $\chi^2/\textrm{dof}$, correlations are taken into account to some extent (cf.\ also \cite{Michael:1994sz}, where in a similar context it has been demonstrated numerically that correlated and uncorrelated $\chi^2$ minimization yield essentially identical results). Since we have only 2 [or 3] degrees of freedom (5 separations, 3 [or 2] fit parameters), not only $\chi^2/\textrm{dof} \ll 1$ but also $\chi^2 \ltapprox 1$, which indicates consistency of the lattice data and our ansatz (\ref{ansatz}), even though we are currently not able, to determine a correlated $\chi^2/\textrm{dof}$. Moreover, note that the resulting values for $\alpha$, $d$ and $p$ are in agreement with phenomenological expectations.


\begin{table}[h]
\caption{\label{fit AEK}$\chi^2$ minimizing fit results of the ansatz (\ref{ansatz}) to the lattice static antiquark-antiquark potential; fitting range $2 \leq r/a \leq 6$; lattice spacing $a \approx 0.079 \, \textrm{fm}$}
\begin{ruledtabular}
\begin{tabular}{ccccc}
channel & $\alpha$ & $d/a$ & $p$ & $\chi^2 / \textrm{dof}$ \\
\hline
scalar isosinglet  &  $0.293(33)$  &  $4.51(54)$  &  $2.74(1.20)$  &  $0.35$ \\
vector isotriplet  &  $0.201(77)$  &  $2.48(69)$  &  $2.0$ (fixed) &  $0.06$ \\
\end{tabular}
\end{ruledtabular}
\end{table}



\section{Heavy antiquark-antiquark binding, existence of tetraquarks}

In Fig.\ \ref{FIG002} it is clear that, if the two $\bar Q$ would be arbitrarily heavy, they would also go arbitrarily deep into the Coulomb potential. In this limit tetraquarks would have an arbitrarily large binding energy both in the scalar isosinglet and in the vector isotriplet channel. However, the heavy $\bar Q$ have a finite mass and the question we now address is, whether the heavy quark mass is large enough, to bind our class of tetraquarks.


\subsection{\label{SEC001}The antiquark-antiquark Hamiltonian}

The potential of Eq.\ (\ref{ansatz}) with the fit parameters from Table~\ref{fit AEK} corresponds to the energy of a static-static-light-light four quark system minus the energy of a pair of static-light ground state mesons. To obtain the energy of a heavy-heavy-light-light system, where the antiquarks have a heavy, but finite mass, one also needs to consider a kinetic term for the heavy antiquarks resulting in the Hamiltonian
\be
\label{EQN001} H = {\mathbf{p}^2 \over 2 \mu} + 2 m_B + V(r) ,
\ee
where $\mu$ is the reduced antiquark mass.

Notice that, because of screening, at large separations each heavy antiquark carries the mass of a $B$ meson and thus $\mu = m_B /2$, whereas at small separations it carries just the energy of a heavy quark $\mu = m_b / 2$. When investigating the existence of four quark bound states, we consider both mass values, which differ by around 5\% ($m_B = 5279 \, \textrm{MeV}$ \cite{PDG}, $m_b = 4977 \, \textrm{MeV}$ in quark models \cite{Godfrey:1985xj}). Another possible source of systematic error is associated with the physical value of the lattice spacing. This error is introduced, when converting the ``size parameter'' $d$ from dimensionless lattice units to physical units. We investigate the magnitude of this error, by using values for the lattice spacing determined by rather different scale setting procedures: $a = 0.079 \, \textrm{fm}$ is used in many ETMC publications and is obtained from $m_\pi$, $f_\pi$ and chiral perturbation theory \cite{Baron:2009wt}, while $a = 0.096 \, \textrm{fm}$ corresponds to identifying the lattice result for the Sommer parameter $r_0$ with its typical value $0.5 \, \textrm{fm}$.


\subsection{An analytical rule for heavy antiquark-antiquark binding}

To get an analytical qualitative understanding, we first derive an approximate rule for the existence/non-existence of a bound state applying the Bohr-Sommerfeld relation corrected by the WKB approximation, to include the zero point energy of $1/2$. For the radial equation in three dimensions we get,
\be
4 \int_{c_1}^{c_2} dr \, \sqrt{2 \mu \bigg(E - V(r) - {l (l+1) \over 2 \mu r^2}\bigg)} = 2 \pi \bigg(n + {3 \over 2}\bigg) ,
\ee
where $c_1$ and $c_2$ are the classical turning points. Inserting (\ref{ansatz}), specializing to angular momentum $l = 0$ and integrating, yields the condition for having at least one bound state,
\be
\mu \alpha d \geq {9 \pi^2 \over 128 \times 2^{1/p} \Gamma^2(1 + 1/2 p)} .
\label{condition}
\ee
The right hand side of Eq. (\ref{condition}) has a rather moderate dependence on the exponent $p$. For example, when $p$ increases from the expected values of $1.0$ to $2.0$, the right hand side only changes from $0.44$ to $0.60$. Thus the existence of a bound state mainly depends on the product of parameters $\mu \alpha d$. With the fit parameters for the scalar isosinglet from Table~\ref{fit AEK} and a bottom quark one roughly obtains $\mu \alpha d \approx 2.5 \, \textrm{GeV} \times 0.29 \times 4.5 \times 0.079 \, \textrm{fm} \approx 1.3 \gg 0.44 \ldots 0.60$, which is a strong indication for the existence of a bound state. A similar calculation for the vector isotriplet yields $\mu \alpha d \approx 0.5$, i.e.\ the existence of a bound state in this channel is rather questionable.


\subsection{Numerical solution of Schr\"odinger's equation}

Note, that the analytical estimates obtained in the previous subsection are rather crude, because e.g.\ the WKB approximation is questionable, when the potential is divergent at the origin as for a Coulomb-like potential. To investigate the existence of a bound state rigorously, we numerically solve the Schr\"odinger equation with the Hamiltonian (\ref{EQN001}). The strongest binding is expected in an s-wave, for which the radial equation is 
\be
\label{EQN002} \left[-{\hbar^2 \over 2 \mu}{d^2 \over dr^2} + 2 m_B + V(r) \right] R(r) = E R(r)
\ee
with the wave function $\psi = \psi(r) = R(r) / r$. We impose Dirichlet boundary conditions $R(r_\textrm{max}) = 0$ at sufficiently large $r_\textrm{max}$ (we checked that results are stable for $r_\textrm{max} \gtapprox 10 \, \textrm{fm}$). The radial equation (\ref{EQN002}) can be solved by standard methods (e.g.\ 4th order Runge-Kutta shooting) up to arbitrary numerical precision.

\begin{figure}[htb]
\includegraphics[scale=1.0]{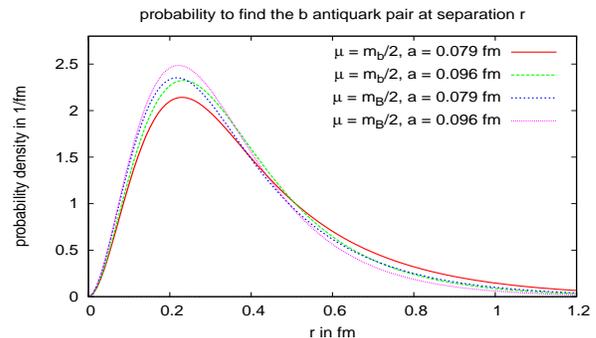}
\caption{\label{FIG003}(Color online). Radial probability density $\propto |R(r)|^2$ for the separation of the heavy antiquarks (scalar isosinglet).}
\end{figure}

In accordance with our analytical estimates we find binding for heavy bottom-bottom tetraquarks in the scalar isosinglet channel. The binding energy $E - 2 m_B \approx 30 \, \textrm{MeV} \ldots 57 \, \textrm{MeV}$ depends to some extent on the reduced mass (either $\mu = m_b/2$ or $\mu = m_B/2$) and on the value of the lattice spacing (either $a = 0.079 \, \textrm{fm}$ or $a = 0.096 \, \textrm{fm}$).
To compute the statistical error of $E - 2 m_B$, we do a Jackknife analysis using the same bins and reduced samples as in section~\ref{SEC005}, when determining the parameters $\alpha$, $d$ and $p$ of the potential ansatz $V(r)$ via $\chi^2$ minimizing fits. Consequently, possibly existing correlations on the level of the lattice correlation functions are partly entering our final results for $E - 2 m_B$ (as already discussed in section~\ref{SEC005}). These results are collected in Table~\ref{binding} together with the ``confidence for binding'', which is just the probability of a negative value for $E - 2 m_B$ assuming a Gaussian distribution for the statistical error. Depending on the concrete choice for the reduced mass $\mu$ and the lattice spacing $a$ the binding energy is negative by around $1.76 \sigma \ldots 3.00 \sigma$. This confirms the existence of an heavy-heavy-light-light tetraquark with a confidence level of $\gtapprox 96 \%$. The probability to find the two heavy antiquarks at separation $r$ is proportional to $|R(r)|^2$ and shown in Fig.\ \ref{FIG003}. The average separation is around $0.25 \, \textrm{fm}$.


\begin{table}[htb]
\caption{\label{binding}Binding energy and confidence for the existence of a heavy-heavy-light-light tetraquark for the scalar isosinglet channel.}
\begin{ruledtabular}
\begin{tabular}{cccc}
$\mu$  &  $a$  &  $E - 2 m_B$  &  confidence for binding \\
\hline
$m_b / 2$  &  $0.079 \, \textrm{fm}$  &  $-30(17) \, \textrm{MeV}$  &  $1.76 \, \sigma$, $\phantom{0}96 \%$ binding \\
$m_b / 2$  &  $0.096 \, \textrm{fm}$  &  $-49(17) \, \textrm{MeV}$  &  $2.88 \, \sigma$, $100 \%$ binding \\
$m_B / 2$  &  $0.079 \, \textrm{fm}$  &  $-38(18) \, \textrm{MeV}$  &  $2.11 \, \sigma$, $\phantom{0}98 \%$ binding \\
$m_B / 2$  &  $0.096 \, \textrm{fm}$  &  $-57(19) \, \textrm{MeV}$  &  $3.00 \, \sigma$, $100 \%$ binding
\end{tabular}
\end{ruledtabular}
\end{table}


Solving Schr\"odinger's equation (\ref{EQN002}) for the vector isotriplet gives strong indication that the potential of this channel is too weak to generate a bound state, i.e.\ a tetraquark. The resulting wave function is essentially a plane wave with positive energy $E - 2 m_B$ within more than $10 \, \sigma$. 


\subsection{Discussion of systematic errors}

Possible sources of systematic error include the concrete choice of values for the $b$ quark mass $m_b$ and the lattice spacing $a$. This has already been addressed in the previous subsection and included in the final results in a rather conservative way.

Moreover, the quality of the lattice results for the heavy antiquark-antiquark potential is not sufficient, to determine the parameters $\alpha$, $d$ and $p$ of the potential ansatz by minimizing a correlated $\chi^2$. We mimimize an uncorrelated $\chi^2$ instead. Our statistical analysis of the binding energy $E - 2 m_B$ (the central quantity studied in this work), however, is a single stringent Jackknife analysis starting already on the level of the lattice correlation functions. As discussed in section~\ref{SEC005} such an analysis partly accounts for potentially existing correlations in $t$ and in $r$. A possibly remaining residual systematic error is expected to be small and to not alter the strong quantitative result (existence of an heavy-heavy-light-light tetraquark with a confidence level of $\gtapprox 96 \%$) in a qualitative way.

Further possible systematic errors are not expected to weaken the binding.
\begin{itemize}
\item The static approximation of a ground state with bottom antiquarks is valid, since the bottom quark mass is a very hard scale compared with the scale $\Lambda_\textrm{QCD}$. Moreover, lattice computations of $B$ mesons \cite{Jansen:2008si,Michael:2010aa} and $b$ baryons \cite{Wagner:2011fs} within the same lattice setup showed that static quarks are a rather good approximation of bottom quarks.

\item In lattice QCD finite volume effects are typically suppressed exponentially. Quantitatively this suppression depends on the extension of the periodic spatial volume $L$ and the mass of the lightest particle, the pion, $m_\pi$ and is proportional to $\exp(-m_\pi L)$. Even for rather simple quantities which can be computed very precisely, $m_\pi L \gtapprox 3 \ldots 4$ is usually a sufficient condition for the finite volume effects to be negligible compared to statistical errors. Therefore, for our results (where $m_\pi L = 3.3$), which exhibit sizeable statistical errors, we do not expect that finite volume effects play an important role. Moreover, the spatial extension of the lattice $L \approx 1.9 \, \textrm{fm}$ seems large compared to the typical size of the tetraquark, which is related to $d \approx 4.51 \, a \approx 0.36 \, \textrm{fm}$ and the average separation of the heavy antiquarks, which is $\approx 0.25 \, \textrm{fm}$. 

\item In what concerns long range forces our Yukawa-like potential falls faster than the original one-pion-exchange Yukawa potential (OPEP). This happens, because our long range part of the potential has too much noise, to measure the small OPEP. We are only sensitive to the dominant exponential  mode in our potential, ie the one due to screening. Notice pions can contribute to the interaction between our two light quarks, and can as well contribute to the pseudoscalar-vector $B B^\ast$ or vector-vector $B^\ast B^\ast$ interaction. In principle the OPEP should enhance the binding at least in some of the channels as it happens for the deuteron.

\item Finally, the light $u/d$ quark masses in the lattice computation are unphysically heavy (corresponding to $m_\pi \approx 340 \, \textrm{MeV}$). Decreasing the light quark masses to their physical values should increase the light cloud radius of a heavy-light meson and, therefore, lead to stronger binding (cf.\ Fig.\ \ref{strings_screened}(a)).
\end{itemize}

Nevertheless, the listed systematic errors should be investigated numerically in the future, to determine their exact effect on the bound state we predict.


\section{Conclusions}

To summarize, we find strong indication for the existence of an antibottom-antibottom-light-light tetraquark bound state. This result is very promising with respect to further lattice QCD calculations with higher statistics or even lighter dynamical quarks, and for experimental searches of doubly bottom hadrons.

It would be most interesting, to extend the present investigation, to study tetraquark resonances such as the BELLE candidate including a $b$ quark and a $\bar b$ antiquark.


\acknowledgements

P.B.\ thanks the hospitality of IFT and FCT grants CERN/FP/116383/2010, CERN/FP/123612/2011.

M.W.\ acknowledges support by the Emmy Noether Programme of the DFG (German Research Foundation), grant WA 3000/1-1.

This work was supported in part by the Helmholtz International Center for FAIR within the framework of the LOEWE program launched by the State of Hesse.



\end{document}